\documentclass{article}
\usepackage{amsmath,amssymb}
\usepackage[english]{babel}

\usepackage[letterpaper,top=2cm,bottom=2cm,left=3cm,right=3cm,marginparwidth=1.75cm]{geometry}

\usepackage{amsmath}
\usepackage{graphicx}
\usepackage[colorlinks=true, allcolors=blue]{hyperref}

\usepackage{hyperref}
\usepackage{setspace} 
\usepackage{lineno}

\hypersetup{
    bookmarks=true,         
    unicode=false,          
    pdftoolbar=true,        
    pdfmenubar=true,        
    pdffitwindow=false,     
    pdfstartview={FitH},    
    pdftitle={My title},    
    pdfauthor={Author},     
    pdfsubject={Subject},   
    pdfcreator={Creator},   
    pdfproducer={Producer}, 
    pdfkeywords={keywords}, 
    pdfnewwindow=true,      
    colorlinks=true,       
    linkcolor=blue,          
    citecolor=blue,        
    filecolor=magenta,      
    urlcolor=cyan           
}

\title{Evolutionary game theory: the mathematics of evolution and collective behaviours}
\author{The Anh Han, Teesside University, UK}

\begin{document}
\maketitle

\begin{abstract}
This brief discusses evolutionary game theory as a powerful and unified mathematical tool to study evolution of collective behaviours. 
It  summarises some of my recent research directions using evolutionary game theory methods, which include i) the analysis of  statistical properties of the number of (stable) equilibria  in a random evolutionary game, and ii) the modelling of safety behaviours' evolution and the risk posed by advanced Artificial Intelligence technologies in a technology development race. 
Finally, it includes an outlook and some suggestions for future researchers. 
\end{abstract}

\section{Introduction}

The mechanisms of emergence and evolution of collective behaviour in populations of abstract individuals, with diverse behavioral strategies in co-presence, have been undergoing mathematical study via Evolutionary Game Theory \cite{nowak:2006pw}. 
They have attracted sustained attention from fields as diverse as  Biology, Economics, Physics, Computer Science and Social Sciences. 
Their systematic study resorts to simulation techniques, thus enabling the study of aforesaid mechanisms under a variety of conditions, parameters, and alternative virtual games. The theoretical and experimental results have continually been surprising, rewarding, and promising.
The problem of explaining (rigorously) the evolution of cooperative behaviour was described as one of the 125 most important questions faced by Sciences in the  Science journal \cite{pennisi2005did}.

Evolutionary Game Theory (EGT) has become a powerful mathematical framework for the modelling and analysis of complex biological, economic and cultural  systems whenever there is frequency dependent selection -- the fitness of an individual does not only depend on its strategy, but also on the composition of the population in relation with (multiple) other strategies \cite{hofbauer:1998mm,sigmund:2010bo}. The payoff from the games is interpreted as individual fitness, naturally leading to a dynamical approach. 
EGT was originated back in 1973 with John Maynard Smith and George R. Price's formalisation of animal contests, extending classical game theory  \cite{von2007theory} to provide the mathematical criteria that can be used to predict the evolutionary outcomes  of interaction among competing strategies \cite{SP73}.



Since its inception, EGT has  been used widely and successfully to study numerous important and challenging questions faced by many disciplines and societies,  such as:  what are the mechanisms underlying the evolution of cooperative behaviour at various levels of organisation (from genes to human society)  \cite{nowak:2006pw}? How to mitigate existential risks such as those posed by climate change  \cite{santos:2011pn} or advanced Artificial Intelligence (AI) technologies \cite{han2019modelling}? What are the roles of cognition and emotions in behavioural evolution  \cite{HanBook2013}?




Diverse mathematical approaches and methods for analysing EGT models have been developed over the years. These include continuous approaches such as replicator equations which assume large population limits, unusually  requiring analysis of  systems of differential equations \cite{hofbauer:1998mm}. In finite size systems,   stochastic approaches are required and usually approximations methods are needed (e.g. using mean-field analysis). Here computer simulations and methods from statistical physics, such as Monte Carlo simulations, are very useful to analyse highly complex systems where analytical results are hard to achieve, for example, when populations are distributed on complex networks \cite{perc2017statistical}. 

Below we are going to present some examples from my recent works, to demonstrate the diverse applications of EGT methods to understand behavioural and biological evolution. 

\section{Applying random polynomial theory to analyse equilibrium properties in random evolutionary games}
Random evolutionary games, in which the payoff entries are random variables, have been employed extensively to model social and biological systems in which  very limited information is available, or where the environment changes so rapidly and frequently that one cannot describe the payoffs of their inhabitants' interactions  \cite{HTG12,DuongHanJMB2016,gokhale:2010pn,DuongTranHanJMB}. 
Equilibrium points of such evolutionary system are the compositions of strategy frequencies where all the strategies have the same average fitness. Biologically, they predict the co-existence of different types in a population and the maintenance of polymorphism.


In random games, due to the randomness of the payoff entries, it is essential to study statistical properties of equilibria. How to determine the distribution of internal equilibria in random evolutionary games is an intensely investigated subject with numerous practical ramifications in ecology, population genetics, social sciences, economics and computer science, 
providing essential understanding of complexity in a dynamical system, such as its behavioural, cultural or biological diversity and the maintenance of polymorphism. Properties of equilibrium points, particularly the probability of observing the maximal number of equilibrium points, the attainability and stability of the patterns of evolutionarily stable strategies
have been studied  \cite{gokhale:2010pn,HTG12}. However, as these prior works used a direct approach that consists of solving a system of polynomial equations, the mathematical analysis was mostly restricted to evolutionary games with a small number of players, due to the impossibility of solving general polynomial equations of a high degree (according to Abel–Ruffini's theorem).

Our recent works analyze  random evolutionary games with an arbitrary number of players \cite{DuongHanJMB2016, DuongTranHanJMB}. The key technique that we develop is to connect the number of equilibria in an evolutionary game to the number of real roots of a system of multi-variate random polynomials \cite{EK95}. Assuming that we consider $d$-player $n$-strategy evolutionary games, then the system consists of $n-1$ polynomial equations of degree $d-1$:  
\begin{equation*}
\label{eq: eqn for fitnessy}
 \sum\limits_{\substack{0\leq k_{1}, ..., k_{n-1}\leq d-1,\\  \sum\limits^{n-1}_{i = 1}k_i \leq d-1  }}\beta^i_{k_{1}, ..., k_{n-1} }\begin{pmatrix}
d-1\\
k_{1}, ..., k_{n}
\end{pmatrix} \prod\limits_{i=1}^{n-1}y_{i}^{k_i} = 0,
\end{equation*}
for $i = 1, \dots, n-1$. Here $\beta^{i}_{k_{1}, ..., k_{n-1} }:= \alpha^{i}_{k_{1}, ..., k_{n} } -\alpha^{n}_{k_{1}, ..., k_{n} }$ where $\alpha^{i_0}_{k_{1}, ..., k_{n} } := \alpha^{i_0}_{i_1,\ldots,i_{d-1}}$ is the payoff of the focal player and $k_i$, $1 \leq i \leq n$, with $\sum^{n}_{i = 1}k_i = d-1$, is the number of players using strategy $i$ in $\{i_1,\ldots,i_{d-1}\}$.
In  \cite{DuongHanJMB2016}, we analyze the mean number $E(n, d)$ and the expected density $f(n,d)$ of internal equilibria in a general $d$-player $n$-strategy evolutionary game when the individuals' payoffs are \textit{independent, normally distributed}.  We provide  computationally implementable formulas of these quantities for the general case and characterize their asymptotic behaviour for the two-strategy games (i.e. $E(2,d)$ and $f(2,d)$), estimating their  lower and upper bounds as $d$ increases. For instance, under certain assumptions on the payoffs, we obtain 
\begin{itemize}
\item Asymptotic behaviour of $E(2,d)$:
\begin{equation*}
\sqrt{d-1}\lesssim E(2,d)\lesssim \sqrt{d-1}\ln(d-1).
\end{equation*}
As a consequence, 
\begin{equation*}
\lim\limits_{d\rightarrow\infty}\frac{\ln E(2,d)}{\ln(d-1)}=\frac{1}{2}.
\end{equation*}
\item Explicit formula of $E(n,2)$:
$
E(n,2)=\frac{1}{2^{n-1}}.
$
\end{itemize}
For a general $d$-player $n$-strategy game, as supported by extensive numerical results, we describe a conjecture regarding the asymptotic behaviours of $E(n,d)$ and $f(n,d)$. We also show that the probability of seeing the maximal possible number of equilibria tends to zero when $d$ or $n$ respectively goes to infinity and that the expected number of stable equilibria is bounded within a certain interval. 

In \cite{DuongTranHanDGA}, we generalize our analysis for random  evolutionary games where the payoff matrix entries are \textit{correlated} random variables. 
In social and biological contexts, correlations may arise in various scenarios particularly when there are environmental randomness and interaction uncertainty such as  when individual contributions are correlated to the surrounding contexts (e.g. due to limited resource). We establish a closed formula for the mean numbers of internal (stable) equilibria and characterize the asymptotic behaviour of this important quantity for large group sizes and study the effect of the correlation. The results show that decreasing the correlation among payoffs (namely, of a strategist for different group compositions) leads to larger mean numbers of (stable) equilibrium points, suggesting that the system or population behavioral diversity can be promoted by increasing independence of the payoff entries.

As a further development, in \cite{DuongTranHanJMB} we derive a closed formula for the distribution of internal equilibria, for both normal and uniform distributions of the game payoff entries. We also provide several universal upper and lower bound estimates,  which are independent of the underlying payoff distribution, for the probability of obtaining a certain number of internal equilibria. The distribution of equilibria provides more elaborate  information about the level of complexity or the number of different states of biodiversity that will occur in a dynamical  system, compared to what obtained with the expected number of internal equilibria.

In short, by connecting EGT to random polynomial theory, we have achieved new results on the expected number and distribution of internal equilibria in multi-player multi-strategy games. Our studies provide new insights into the overall complexity of dynamical systems, including biological and social  ones, as the numbers of players and  strategies in an interaction within the systems increase. As the theory of random polynomials is rich, we expect that our novel approach can be extended to obtain results for other more complex models in population dynamics such as the replicator-mutator equation and evolutionary games with environmental feedback. 

\section{Evolutionary game  modelling of safety behaviours in an Artificial Intelligence development race}
Rapid technological advancements in Artificial Intelligence (AI), together with the growing deployment of AI in new application domains such as robotics, face recognition, self-driving cars, genetics, are generating an anxiety which makes companies, nations and regions think they should respond competitively. AI appears for instance to have instigated a race among chip builders, simply because of the requirements it imposes on the technology.  Governments are furthermore stimulating economic investments in AI research and development as they fear of missing out, resulting in a racing narrative that increases further the anxiety among stake-holders. 

Races for supremacy in a domain through AI may however have detrimental consequences since participants to the race may well ignore ethical and safety checks in order to speed up the development and reach the market first. AI researchers and governance bodies, such as the EU and Future of Life institute, are urging to consider together both the normative and the social impact of major technological advancements concerned. However, given the breadth and depth of AI and its advances, it is not an easy task to assess when and which AI technology in a concrete domain needs to be regulated. This issue was, among others, highlighted in the recent EU White Paper on AI.  Data to estimate the risk of a technology is usually limited, especially at an early stage of its development or deployment. 

Here, we summarise our recent works \cite{han2019modelling,han2020Incentive}  examining this problem theoretically using EGT methods, resorting to a novel innovation dilemma where technologists can choose a safe (SAFE) vs risk-taking (UNSAFE) course of development. Companies race towards the deployment of some AI -based product in some domain X. They can either carefully consider all data and AI pitfalls along the way (SAFE) or else take undue risks by skipping some tests just to speed up the process (UNSAFE). Overall, SAFE are costlier strategies and take more time to implement than UNSAFE strategies, allowing UNSAFE strategists to further claim significant benefits from reaching technological supremacy. 

\begin{figure*}[h!]
\centering
\includegraphics[width=0.95\linewidth]{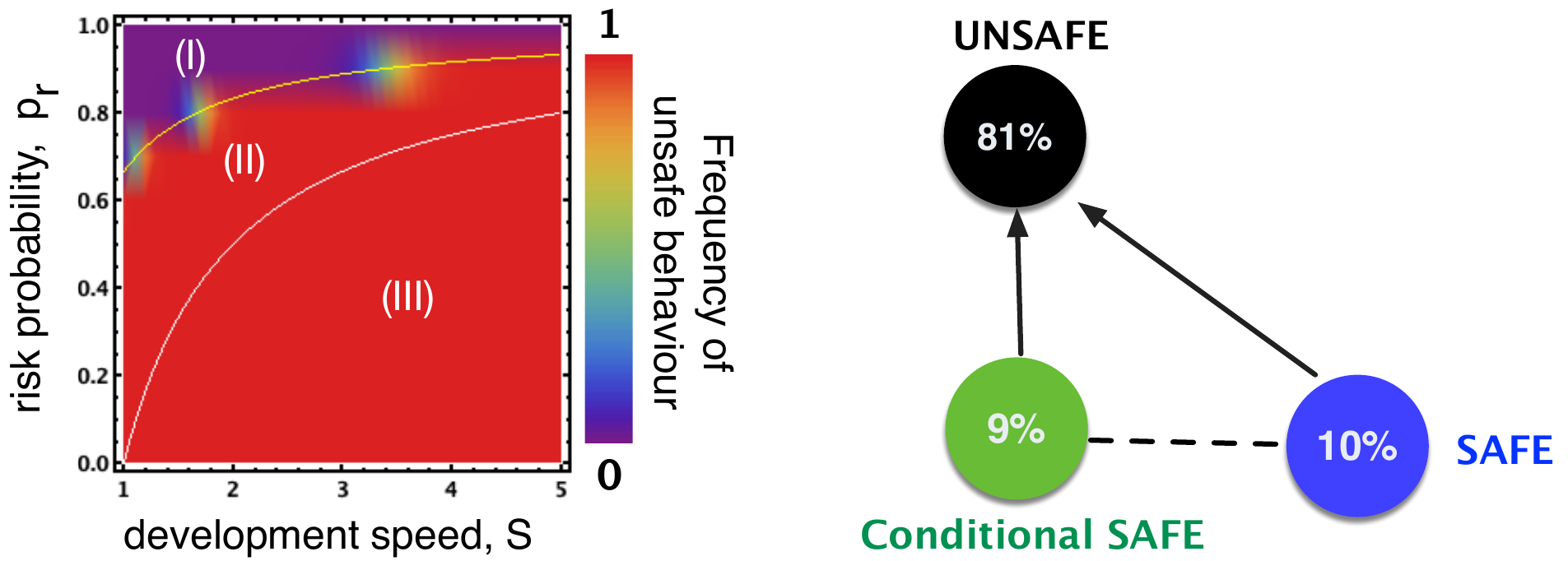} 
\caption{Frequency of unsafe behaviour as a function of development speed and the disaster risk, in absence of incentives (see Reference  \cite{han2019modelling}). In regions (I) and (III), safe and unsafe/innovation, respectively, are the preferred collective outcome and are selected by natural selection, thus no regulation being required. Region (II) requires regulation as safe behaviour is preferred but not selected. The analytical conditions of the boundary lines can be identified for region (II), using EGT methods: $1-1/s < p_r < 1-1/(3s)$. 
}
\label{fig:vary_pr-otherbetas}
\end{figure*}

In more detail, we posit that it requires time to reach domain supremacy through AI (DSAI), modelling this by a number of development steps or technological advancement rounds \cite{han2019modelling}. In each round the development teams (or players) need to choose between one of two strategic options: to follow safety precautions (the SAFE action) or ignore safety precautions (the UNSAFE action). Moreover, to avoid being exploited, we also consider a conditional SAFE strategy that plays SAFE only if the others in the competition did so in the past. 

Because it takes more time and more effort to comply with precautionary requirements, playing SAFE is not just costlier, but implies slower development speed too, compared to playing UNSAFE. We consequently assume that to play SAFE involves paying a cost $c>0$, while playing UNSAFE costs nothing ($c=0$). Moreover, the development speed of playing UNSAFE is $s>1$ whilst the speed of playing SAFE is normalised to $s=1$. The interaction is iterated until one or more teams establish DSAI, which occurs probabilistically, i.e. the model assumes, upon completion of each round, that there is a probability $\omega$ that another development round is required to reach DSAI—which results in an average number $W=(1-\omega)^{-1}$ of rounds per competition/race. We thus do not make any assumption about the time required to reach DSAI in a given domain. Yet once the race ends, a large benefit or prize $B$ is acquired that is shared amongst those reaching the target simultaneously.

We pitch all development teams (e.g. AI companies, nations) together in the time evolution of the adoption of SAFE,  UNSAFE or conditional SAFE strategies. We resort to a  EGT stochastic method for finite populations \cite{nowak:2004pw}. For example, in Figure \ref{fig:vary_pr-otherbetas} we consider a population of 100 development teams who can at first randomly choose one of the three strategies. Briefly, we need to calculate a stationary distribution of a Markov chain whose nodes represent strategies being considered (i.e. three nodes in our case, see the right panel), and the transitions between the nodes represent the probability that a strategy invades a population where individuals adopt another strategy (see details in \cite{han2019modelling}).


As a result, we identify conditions under which safe or risk-taking behaviour emerges, and when they are collectively preferred, leading to a greater population social welfare. We next explored ways to influence it towards safe and beneficial outcomes, namely when and how to sanction unsafe decisions made by stake-holders or reward compliant ones. Finally, we identify when regulations need to be put in place to favour outcomes most beneficial for all, but at the same time taking care to avoid strict regulations that would be introduced too early and thereby stifle innovation \cite{han2019modelling,han2020Incentive}.



\subsection*{Lessons for AI governance policies}

We find that the time-scale in which domination or supremacy in an AI domain can be achieved plays a crucial role in determining when regulatory actions are required \cite{han2019modelling}. For instance, it would probably take very long until we have an AI capable of doing anything that humans do (one usually termed Artificial General Intelligence). Still, in  many domains, such as chess playing, AI already outperforms humans. It would not take very long until self-driving cars become safer than average human drivers.  

We find that, in short-term result scenarios, companies that ignore safety precautions are bound to win in our simulations, and hence they should be regulated. Nonetheless, in this case, the exact requirements of regulations depend on finding a balance between the desirable innovation speed and the risk of its negative externalities.

Differently, in a long-term result scenario, screening for unsafe actions ensures that only when the risk is low will winning companies act in an unsafe manner. Such risk-taking, as opposed to compliance with safety measures, should be regulated for society's benefit. It goes without saying that, in both time-scales, only when individual benefits conflict with the overall societal interests, will explicit regulation of unsafe actions become paramount. 

These findings indicate that, when defining codes of conduct and regulatory policies for AI, first of all, a clear understanding about the timescale of the race is required for effective AI governance. Regulation might not always be necessary and could even have detrimental effects if not timely applied in the right circumstances. 

Indeed, we explicitly tested in our simulations what would happen if always companies that take risks are sanctioned \cite{han2020Incentive}, reducing their speed but at the cost of speed reduction by the sanctioning party. As anticipated, over-regulation, conducive to beneficial innovation being stifled, occurred whenever the gain from speeding up out-benefited the taking of risk.

Yet an issue remains to be solved for proper regulation: Even if we can assess the game's timescale, we still need to estimate the measures of risk and gain associated with risk-taking behaviours. We need data to do so, but it is usually not yet available at an early stage of development. 

Our latest finding though suggests a way out based on the idea of voluntary safety agreements. That is, desirable outcomes are achievable without any over-regulation whatsoever if companies have the freedom of choice between independently pursuing their course of actions or else establishing instead binding agreements to act safely. Sanctioning can then be applied only against those that do not abide by their commitment pledges. 
Thus, our analysis indicates the  need  to facilitate this option, enabling AI companies to voluntarily commit to safety agreements without repercussion should they choose to opt out.


\section{Summary and outlook}
Since its inception, research in EGT has been both exciting and challenging, as well as highly rewarding and inspiring. 
On the one hand, analysing large-scale dynamical systems of interactions among several parties, which is the primary goal of EGT,  is highly complex and requires innovative combinations of diverse mathematical methods and simulation techniques.  
On the other hand, EGT research has proven extremely powerful and found its applications in so many fields, leading to important findings  often being  reported in prestigious venues.    

Despite the many years of active research, there are still a  number of important open problems in EGT research. The problem of explaining the mechanisms of collective behaviour such as cooperation and altruism, is still far from settled \cite{pennisi2005did}. Also,  extending and generalising the existing set of mathematical techniques and modelling  tools to capture and understand realistic and ever more complex systems, is crucial. For example, modern societies become increasingly more convoluted with the advancement of technologies, changing the way  humans  live and interact with others. How EGT  can be applied to model this hybrid society and understand its dynamics is very challenging. But solving it can prove very rewarding as it can provide insights to design appropriate mechanisms to ensure the greatest benefit for our societies, or at least to avoid existential risks that we  might otherwise have to face. 

Researchers and students with background in applied and pure mathematics, with an   interest in interdisciplinary research, complex systems and human behaviors, would be well equipped to tackle these research challenges. The reward would also be huge, both in terms of research enjoyment and  outcomes. 



\bibliographystyle{alpha}
\bibliography{sample}

\end{document}